\documentclass[twocolumns,bibyear]{aa} 
\usepackage{graphicx,epstopdf}
\usepackage{url}
\usepackage{txfonts}
\usepackage{amsmath}
\def\Q{\ifmmode\mathcal{Q}\else$\mathcal{Q}$\fi}

\newcommand{\RNum}[1]{\uppercase\expandafter{\romannumeral #1\relax}}

\usepackage{color}

\begin{document} 

   \title{Assessing the accuracy of star formation rate measurements by direct star count in molecular clouds}

   \author{Sami Dib\inst{1}, Jian-Wen Zhou\inst{2}, S\'{e}bastien Comer\'{o}n\inst{3,4}, Luis E. Gardu\~{n}o\inst{5}, Valery V. Kravtsov\inst{6}, Paul C. Clark\inst{7}, Guang-Xing Li\inst{8}, Maritza A. Lara-L\'{o}pez\inst{9}, Tie Liu\inst{10}, Mohsen Shadmehri\inst{11}, James R. Doughty\inst{7}}
 
   \institute{Max Planck Institute for Astronomy, K\"{o}nigstuhl 17, 69117, Heidelberg, Germany\\
      \email{sami.dib@gmail.com; dib@mpia.de}
    \and Max Planck Institute for Radioastronomy, Auf dem H\"{u}gel 69, 53121 Bonn, Germany
    \and Departamento de Astrof\'{i}sica, Universidad de La Laguna, 38200, La Laguna, Tenerife, Spain 
    \and Instituto de Astrof\'{i}sica de Canarias, 38205, La Laguna, Tenerife, Spain
     \and Instituto Nacional de Astrof\'{i}sica, \'{O}ptica y Electr\'{o}nica (INAOE), Luis Enrique Erro No.1, Tonantzintla, Pue., C.P. 72840, M\'{e}xico
    \and Sternberg Astronomical Institute, Lomonosov Moscow State University, University Avenue 13, 119899, Moscow, Russia
    \and School of Physics and Astronomy, Cardiff University, Queen's Buildings, The Parade, Cardiff CF24 3AA, United Kingdom
    \and South-Western Institute for Astronomy Research, Yunnan University, Yunnan, Kunming 650500, People's Republic of China
    \and Departamento de F\'{i}sica de la Tierra y Astrof\'{i}sica, Instituto de F\'{i}sica de Part\'{i}culas y del Cosmos, IPARCOS. Universidad Complutense de Madrid (UCM), E-28040, Madrid, Spain
    \and Shanghai Astronomical Observatory, Chinese Academy of Sciences, 80 Nandan Road, Shanghai 200030, People's Republic of
China
    \and Department of Physics, Faculty of Sciences, Golestan University, Gorgan 49138-15739, Iran
      }                 
          
\authorrunning{Dib et al.}
\titlerunning{SFR from direct star count}
         
 
\abstract{Star formation estimates based on the counting of young stellar objects (YSOs) are commonly carried out for nearby star-forming regions in the Galaxy, and in principle could be extended to any star-forming region where direct star counts are possible. With this method, the SFRs are measured using the counts of YSOs in a particular class, a typical mass, and the lifetime associated with this class. Another variant of this method is to use the total number of YSOs found in a star-forming region along with a characteristic YSO timescale. However, the assumptions underlying the validity of this method, such as that of a constant star formation history (SFH), have never been fully tested, and it remains unclear as to whether or not the method is valid for all protostellar classes. In this work, we use Monte Carlo models to test the validity and robustness of the method. We build synthetic clusters in which stars form at times that are randomly drawn from a specified SFH distribution function. The latter is either constant or time dependent, with a burst like behavior. The masses of the YSOs are randomly drawn from a stellar initial mass function (IMF), which can be either similar to that of the Milky Way field or be variable within the limits of the variations observed among young stellar clusters in the Galaxy. For each star in every cluster, the lifetimes associated with the different protostellar classes are also randomly drawn from Gaussian distribution functions centered around their most likely value as suggested by the observations. We find that only the SFR derived using the Class 0 population can reproduce the true SFR at all epochs, and this is true irrespective of the shape of the SFH. For a constant SFH, the SFR derived using the more evolved populations of YSOs (Class I, Class F, Class II, and Class III) reproduce the real SFR only at later epochs, which correspond to epochs at which their numbers have reached a steady state. For a time-dependent burst-like SFH, all SFR estimates based on the number counts of the evolved populations fail to reproduce the true SFR. We show that these conclusions are independent of the IMF. We argue that the SFR based on the Class 0 alone can yield reliable estimates of the SFR. We also show how the offsets between Class I- and Class II-based SFRs and the true SFR plotted as a function of the number ratios of Class I and Class II versus Class III YSOs can be used in order to provide information on the SFH of observed molecular clouds.}


 \maketitle

%

\section{Introduction and motivation}\label{intro}
 
Understanding the rate at which molecular clouds convert their gas reservoirs into stars is of particular importance for many branches of astrophysics. The star formation rate (SFR) determines a wide range of the observed properties of galaxies and affects their chemical and dynamical evolution (Helou et al. 2000; Dib et al. 2009; Maraston et al. 2010; Dib 2011; Dariush et al. 2016; Pacifici et al. 2016; Barrera-Ballesteros et al. 2021; Grisdale et al. 2022;  Gardu{\~n}o et al. 2023; Egorov et al. 2023). Within individual molecular clouds, the SFR is regulated by a number of physical processes that include the clouds' self-gravity, supersonic turbulence and the way it is driven, their chemical composition, the incident radiation field, and tidal fields (Krumholz \& McKee 2005; Dib et al. 2007,2008,2010; Padoan \& Nordlund 2011; Glover \& Clark 2012; Federrath \& Klessen 2012; Dib et al. 2020; Schneider et al. 2022; Rani et al. 2022; Dib 2023a, Li 2024; Zhou et al. 2024a,b). On large scales, the SFR can be regulated by other processes, such as the gravity due to the existing stellar populations (Dib et al. 2017a; Shi et al. 2018; Marchuk 2018; Pessa et al. 2022), galactic shear (Seigar et al. 2005; Dib et al. 2012; Aouad et al. 2020), and the overall galactic potential (Jog 2014; Meidt et al. 2018). 

The SFR can be measured using a variety of star-formation indicators. Some of these methods rely on the stellar light that is directly emitted by young stars, such as in the ultraviolet, or by dust-processed stellar light in the infrared. For a detail review of these methods, we refer the reader to Calzetti (2013). While it should be noted that these methods are mostly employed for external galaxies where stellar populations are not well resolved, they make assumptions about the shape of the stellar IMF, and commonly assume that it is universal within and across galaxies. Recent work has shown that this is not the case, either on individual molecular clouds scales (Dib 2014; Dib et al. 2017b; Dib 2023b) or for entire galaxies (Dib \& Basu 2018; Dib 2022). In nearby star forming regions in the Milky Way, a different method is usually used to infer the SFR. This method is based on the direct counting of young stellar objects (YSOs)\footnote{We use the acronym YSO to designate the young stellar population in its ensemble. When referring specifically to young YSOs (Class 0, Class I, and the F Class), we employ the terms protostars. The older YSOs (Class II and Class III) are referred to as pre-main sequence (PMS) stars.}, which are optimally identified in the infrared (Forbrich 2009; Enoch et al. 2009; Evans et al. 2009; Heiderman et al. 2010; Guthermuth et al. 2011; Ybarra et al. 2013; Jose et al. 2013; Strafella et al. 2015; Stutz \& Kainulainen 2015; Jose et al. 2016; Fischer et al. 2016; Hasan et al. 2023; Tobin \& Sheehan 2024). The number of YSOs in a given evolutionary stage in a star-forming region, $N_{\rm, i}$ , is converted into a ${\rm SFR_{i}}$ using 

\begin{equation}
{\rm SFR}_{i} = N_{i}  \frac{\langle M \rangle}{\tau_{i}},
\label{eq1}
\end{equation}

\noindent where the index $i$ runs over the different protostellar classes ($i$=0, I, F, II, and III for objects in Class 0, Class I, Class F, Class II, and Class III, respectively), $\langle M \rangle$ is the mean protostellar mass present in the region, and $\tau_{i}$ is the typical lifetime of the YSOs spent in each class. The accuracy of the SFR measurements using Eq.~\ref{eq1} depends on the validity of some of the assumptions that are adopted when calculating the average mass $\langle M\rangle$ and protostellar lifetimes. For example, for Class II YSOs, $\tau_{\rm II}$ is often taken to be $\approx 2$ Myr. The mean mass is usually assumed to be $\langle M\rangle=0.5$ M$_{\odot}$, which corresponds to the mean mass in a Milky Way-like IMF. These types of simplifications could potentially generate a bias in the estimates of the SFR. Another source of uncertainty that can affect the SFR measurements using YSO counts is the incompleteness of the YSO censuses and the level of incompleteness can vary with the evolutionary stage of the YSOs as well as with the specific details of how the observational data were obtained and reduced (Gutermuth et al. 2009; Winston et al. 2010; Broekhoven-Fiene et al. 2014; Dunham et al. 2015; Gutermuth \& Heyer 2015). In this work, we attempt to estimate whether or not such biases exist and to determine their amplitudes. We approach this problem by generating synthetic populations of $N_{*}$ YSOs that form in a molecular cloud with a prescribed star formation history (SFH). The YSOs have masses that are drawn from an IMF and are assigned protostellar lifetimes that span the realistic range of each class. We then measure the real SFR and compare it to the SFR derived using Eq.~\ref{eq1} for YSOs in different classes. In \S.~\ref{mod}, we describe the elements of the model, namely the protostellar mass function from which stellar masses are drawn, the function that describes the SFH of the YSOs, and their lifetimes spent within each class. In \S.~\ref{res}, we present the results of the time evolution of the SFRs derived from the various protostellar classes and compare them to the real SFR, exploring both the effects of the SFH and of the IMF. In \S.~\ref{correct} we measure the offsets between the real SFRs and those derived using the Class I and Class II populations. We explore the correlation between these offsets and the ratios between the numbers of Class I and II YSOs and those of Class III. Finally, in  \S.~\ref{conc}, we present our conclusions. 
  
\section{The model}\label{mod}

\subsection{The protostellar mass function}

As we do not account for the mass evolution of YSOs, which is minimal for low-mass stars, we also refer to the protostellar mass function as the stellar IMF and assume that it can be described with a tapered power-law function (TPL; Dib 2014; Dib et al. 2017b; Dib 2023b). The TPL is characterized by two power laws in the low- and high-mass regimes, and a characteristic mass (i.e., peak mass) and is given by

\begin{equation}
\phi_{\rm i} (M)= \phi_{\rm 0}  M^{-\Gamma-1} \left\{1-\exp\left[-\left(\frac{M}{M_{\rm ch}}\right)^{\gamma+\Gamma}\right] \right\}\,
\label{eq2}
,\end{equation}

\noindent where $\phi_{0}$ is a normalization term, $\Gamma$ is the exponent of the power law in the high-mass end, $\gamma$ is the exponent in the low-mass regime, and $M_{\rm ch}$ is the characteristic mass. The Salpeter (1955) value for the slope in the high-mass regime corresponds to $\Gamma=1.35$. The Milky Way values for the two other variables for a single-star IMF (i.e., corrected for binarity) are $\gamma=0.51$ and $M_{\rm ch}=0.35$ M$_{\odot}$ (Parravano et al. 2011). In this work, we consider cases where the parameters of the clusters' IMF are assigned Milky Way-like values, and others where we vary the three parameters within the ranges that are permitted by the observations of young Galactic stellar clusters. The minimum and maximum stellar masses we consider are 0.02 M$_{\odot}$ and 50 M$_{\odot}$, respectively.

\begin{figure}
\begin{center}
\includegraphics[width=\columnwidth] {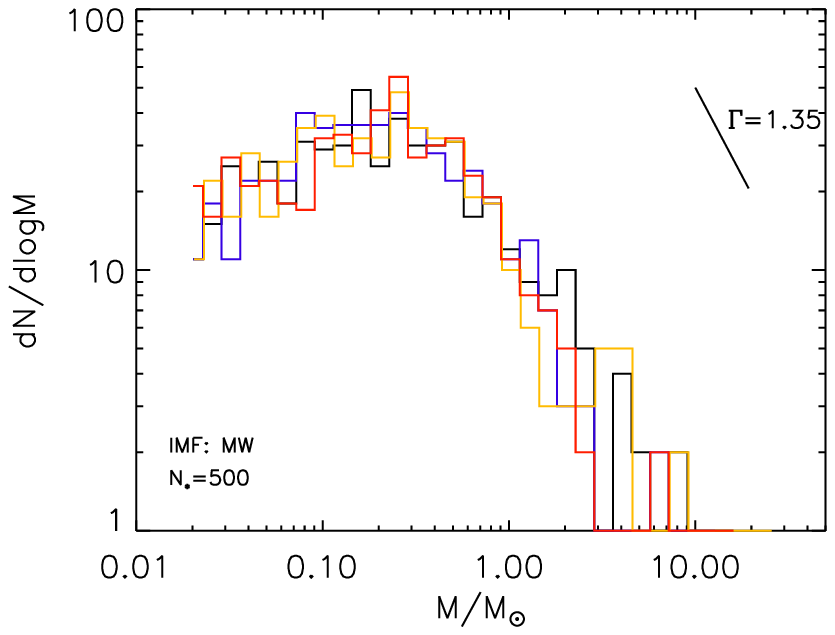}
\caption{Four realizations of the Milky Way-like protostellar mass function. The masses of $N_{*}=500$ protostars are randomly drawn from a Milky Way-like stellar mass function in the stellar mass range of 0.02 to 50 M$_{\odot}$.}
\label{fig1}
\end{center}
\end{figure} 

\begin{figure}
\begin{center}
\includegraphics[width=\columnwidth] {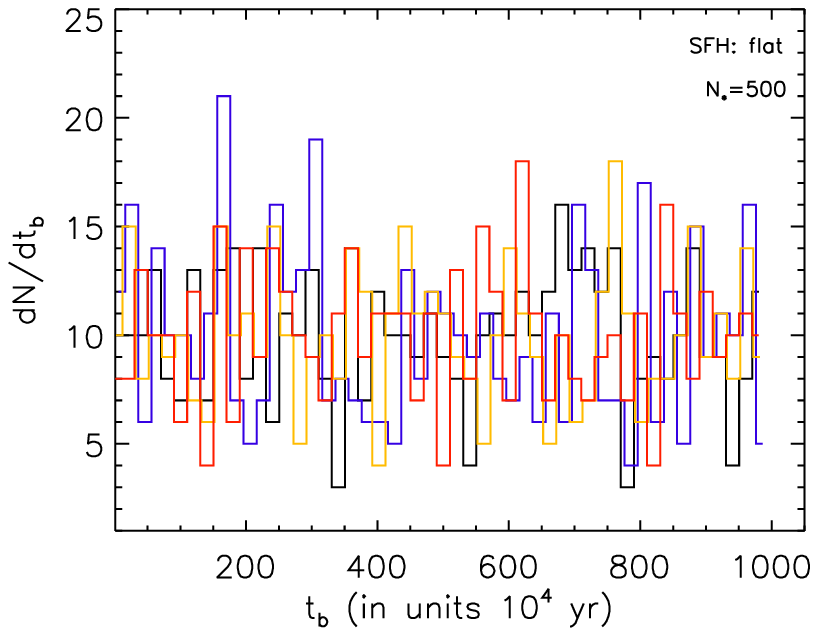}\\
\includegraphics[width=\columnwidth] {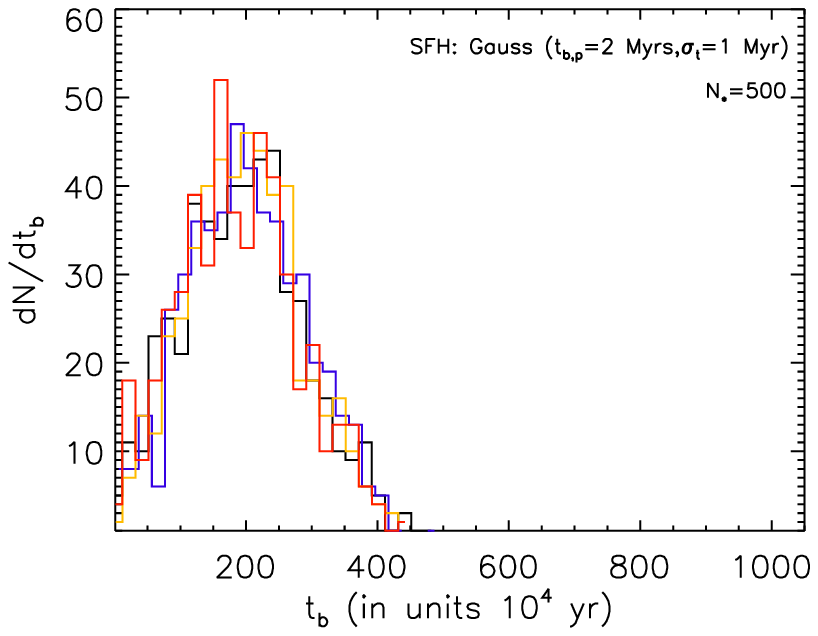}
\caption{Top: Four realizations of the birth time of protostars with a flat, constant SFH. Bottom: Four realizations of the birth time of protostars with a Gaussian SFH. The peak and standard deviation of the Gaussian are set to $t_{\rm b,p}=2$ Myr and $\sigma_{\rm t}=1$ Myr. For improved visibility, the values of $t_{\rm b}$ are binned with a bin size of $2\times10^{5}$ yr.}
\label{fig2}
\end{center}
\end{figure}  

\subsection{Protostellar classes and associated lifetimes} \label{classlifetimes}

The classification into Classes I, II, and III was initially proposed by Lada \& Wilking (1984) and is based on the value of the slope of the spectral energy distribution (SED) in the wavelength range of 2 to 20 $\mu$m. This classification system was subsequently used in numerous studies (Andr\'{e} \& Montmerle 1994; Evans et al. 2009; Rebull et al. 2014; Furlan et al. 2016; Sharma et al. 2017; Pokhrel et al. 2020; Kuhn et al. 2021; Sun et al. 2022). The Class 0 classification was later introduced by Andr\'{e} et al. (2000) and is based on the indirect inference of the presence of a YSO based on the detection of a compact submillimeter source and a collimated CO outflow. These protostellar classes are associated with different phases of the formation of stars and each phase is characterized by a specific shape of the SED. The star-formation processes taking place in these phases are as follows: In Class 0, a YSO is formed in the central region of a protostellar core with an envelope mass that is far in excess of the YSO mass; in Class I, there is collapse of the envelope onto the central object with the transition between Class 0 and Class I being the point in time at which the envelope mass and the mass of the protostar are almost equal; in Class II, a disk emerges around the central star; and in Class III, the disk dissipates by various processes, such as the formation of planets, photo-evaporation, and tidal stripping. An intermediate class between Class 0 and Class I has been proposed by Greene et al. (1994) and was labelled the Flat class (i.e., a relatively flat SED in the wavelength range of 2 to 20 $\mu$m) but it was shown that this class is close to Class I in terms of the evolutionary status of the YSOs as it is associated with signs of a collapsing envelope (Calvet et al. 1994). Recent work shows that genuine flat-spectrum YSOs are rare, once contamination by reddened Class II YSOs and dusty star-forming galaxies is corrected for (Pokhrel et al. 2023)\footnote{The exact nature of the flat-spectrum YSOs is still under investigation. A recent analysis using observations from the ATACAMA Compact Array (ACA) suggests that these objects are more likely disk dominated than envelope dominated and, as such, are more likely to be late-stage protostars or edge-on Class II YSOs (Federman et al. 2023).}

\begin{figure}
\begin{center}
\includegraphics[width=\columnwidth] {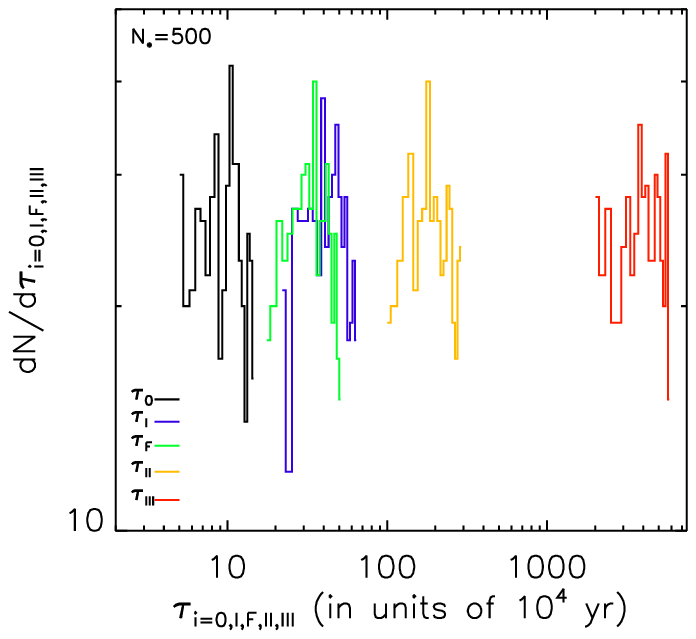}
\caption{One realization of the distribution functions of the lifetimes associated with each protostellar class. The lifetimes are drawn using Eq.~\ref{eq4} with the mean value and standard deviation associated with each lifetime (see text for details). The bin size for each protostellar lifetime distribution is adjusted to the mean value of that  distribution, namely the bin sizes are 5$\times10^{3}$ yr, $2.2\times10^{4}$ yr, $1.75\times10^{4}$ yr, $10^{5}$ yr, and $2\times10^{6}$ yr for the Class 0, Class I, the F Class, Class II, and Class III, respectively.}
\label{fig3}
\end{center}
\end{figure} 

The exact duration of each evolutionary stage for YSO classes is relatively uncertain. Estimates of the mean duration of each of these stages depend on the number of objects found in each class in star-forming regions and this may be affected by misclassifications due for example to YSOs being seen edge-on (Gutermuth et al. 2009). The usual approach is to use Eq.~\ref{eq1} and the numbers of YSOs found in different classes in order to derive relative ages for each class. The assumption is made that YSOs are formed at a constant rate over time and that their formation occurs over timescales that are typically longer than the longest duration of any class. By constraining the age of one of the classes, such as that of the Class II T Tauri stars, which is assumed to be of the order of their disk lifetime \footnote{The disk lifetime of a class II YSO is estimated as being the timescale over which the fraction of YSOs with a disk decays as a function of the mean isochronal ages of star forming regions of a certain range of ages (Haish et al. 2001; Hern\'{a}ndez et al. 2007,2008; Mamajek 2009).}, one can then derive the corresponding duration of the other classes for each YSO. A number of studies measured these timescales in nearby star-forming clouds and found them to fall in the range of $0.005$ to $0.17$ Myr for Class 0 YSOs (Andr\'{e} et al. 1993; Froebrich et al. 2006; Enoch et al. 2009; Fischer et al. 2017; Kristensen \& Dunham 2018), between $0.1$ and $0.7$ for Class I YSOs (Wilking et al. 1989; Greene et al. 1994; Kenyon \& Hartmann 1995; Hatchell et al. 2007), and between $0.4$ and $5$ Myr for Class II YSOs (Wilking et al. 1989; Kenyon \& Hartmann 1995; Hern\'{a}ndez et al. 2008), and found ages in excess of 2 Myr for Class III YSOs (Wilking et al. 2005). 

In the present work, we adopt the values of the lifetimes measured by Evans et al. (2009). These were derived from observations of several nearby molecular clouds in the the {\it Spitzer} legacy project, ``From Molecular Cores to Planet-forming Disks'' (c2d). Evans et al. (2009) found that the mean lifetimes of Class 0, Class I, and the Flat Class are $\bar{\tau}_{0}=0.1$ Myr, $\bar{\tau}_{\rm I}=0.44$ Myr, and $\bar{\tau}_{\rm F}=0.35$ Myr. These were derived assuming a constant SFR and a lifetime for Class II of $\bar{\tau}_{\rm II}=2$ Myr and include a correction of the photometry for extinction effects. The uncertainties on these lifetimes are quite large. For Class II, the $1\sigma$ uncertainty is about $\sigma_{\tau_{\rm II}} =1$ Myr. Given all the assumptions, such as that of a constant SFR, the potential confusion of a fraction of the YSOs with extragalactic sources, and differences in the relative numbers of YSOs found in each region of the c2d cloud sample, we adopt uncertainties on the lifetimes that are $50\%$ of their adopted values, such that $\sigma_{\tau_{0}}=0.05$ Myr, $\sigma_{\tau_{\rm I}}=0.22$ Myr, and $\sigma_{\tau_{\rm F}}=0.175$ Myr. These results were corroborated by the study of Hsieh \& Lai (2013), who reanalyzed the c2d data and found similar results for the lifetimes to those obtained by Evans et al. (2009). For the Class III YSOs, their census in the c2d is incomplete, and therefore their associated lifetimes were not estimated. In general, the lifetime associated with Class III YSOs is uncertain. An approximation of this duration is the Kelvin-Helmotz timescale, which is the typical time during which a young star contracts and radiates its kinetic energy and descends the Hayashi track to reach the main sequence (Hayashi 1961). The Kelvin-Helmotz timescale is given by $\tau_{\rm KH}\approx (G M^{2})/(2 R L)$, where $G$ is the gravitational constant and $M$, $R$, and $L$ are the star's mass, radius, and stellar luminosity, respectively. For solar-mass stars, $\tau_{KH}\approx 30$ Myr, and this value is lower for higher-mass stars and is substantially higher for low-mass stars. However, as we follow the evolution of the young stellar populations for the first $10$ Myr, the choice of any value of $\tau_{KH}$ that is larger than 30 Myr will make little difference to our calculations. Here, we consider a single value across the entire stellar mass range of $\bar{\tau}_{\rm III}=40$ Myr with $\sigma_{\tau_{\rm III}}=20$ Myr.

In the present work, we draw the lifetimes of the population of YSOs from Gaussian distribution functions given by 

\begin{equation}
P(\tau_{i})=\frac{1}{\sigma_{\tau_{i}} \sqrt{2 \pi}} \exp\left(-\frac{1}{2} \left(\frac{\tau_{i}-\bar{\tau}_{i}}{\sigma_{\tau_{i}}}\right)^{2}\right),
\label{eq3}
\end{equation}

\noindent where $i=0$, I, F, II, and III correspond to Class 0, Class I, the F Class, Class II, and Class III, respectively, and $\bar{\tau}_{i}$ and $\sigma_{\tau_{i}}$ are the characteristic lifetimes and the corresponding $1\sigma$ uncertainties for each class taken from Evans et al. (2009). 

\subsection{The star formation history}

\begin{figure*}
\begin{center}
\includegraphics[width=0.8\columnwidth] {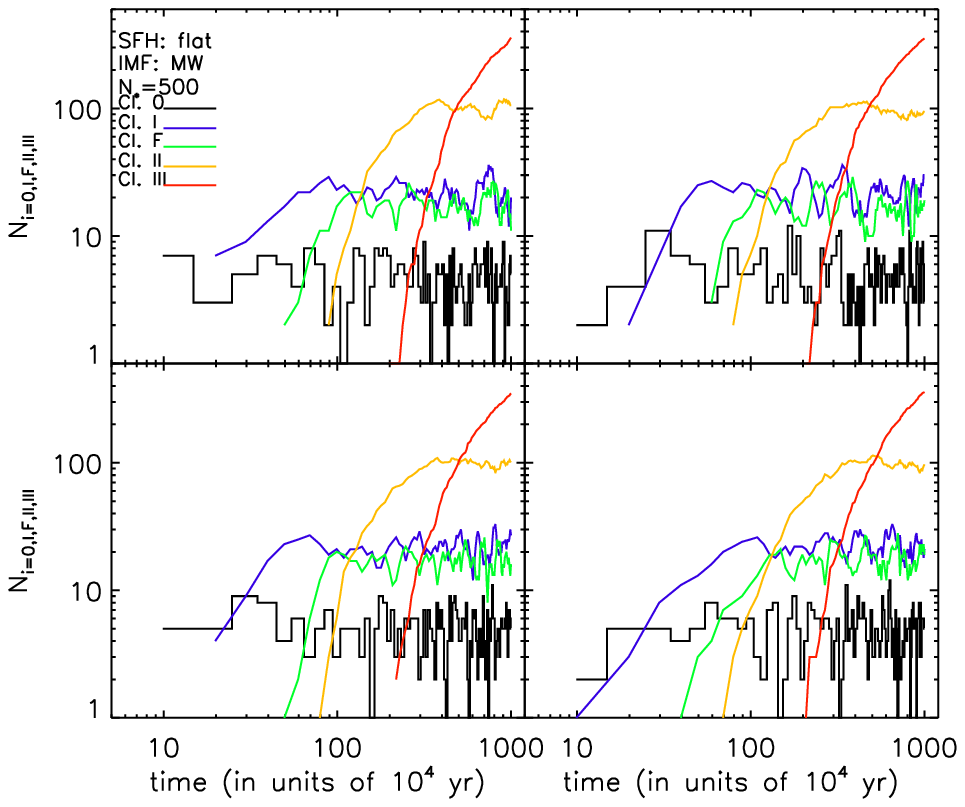}
\hspace{1.5cm}
\includegraphics[width=0.8\columnwidth] {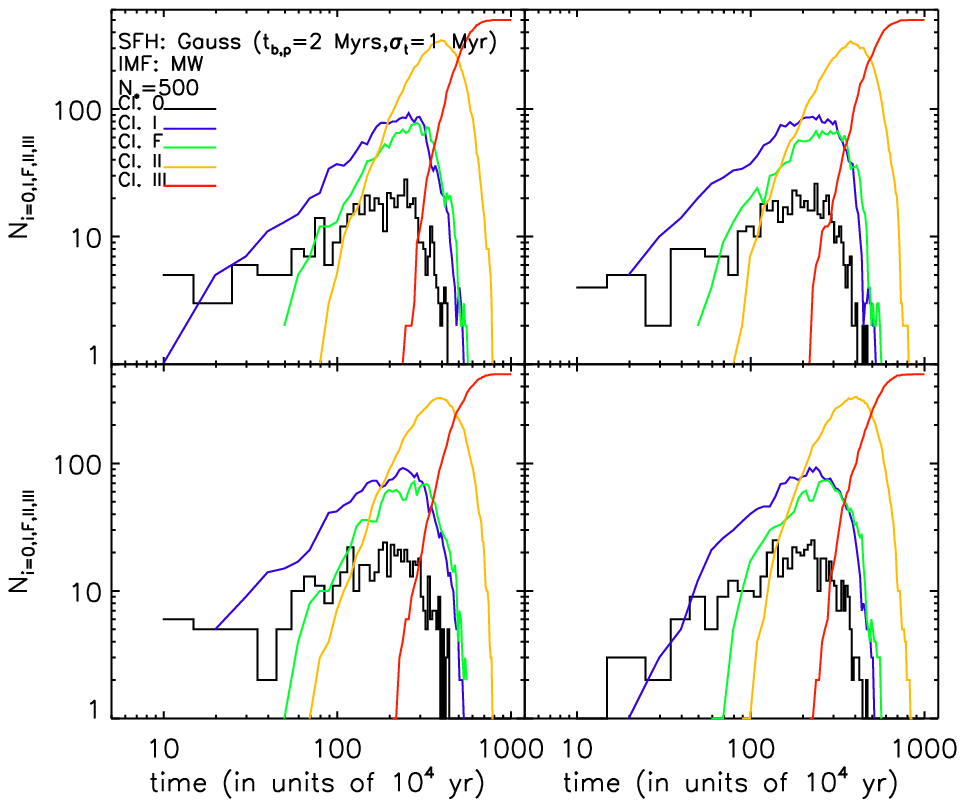}
\vspace{1cm}
\caption{Left: Four realizations of the time evolution of the number of YSOs in different protostellar classes. Each cluster has $N_{*}=500$, a Milky Way-like IMF and a constant SFH. Right: Similar to left panel, but here each cluster has $N_{*}=500$, a Milky Way-like IMF, and a Gaussian-like SFH. The Gaussian's peak is at 2 Myr and the standard deviation is 1 Myr. The bin-size on the x-axis is $10^{5}$ yr and the same value is used in subsequent plots.} 
\label{fig4}
\end{center}
\end{figure*} 

\begin{figure*}
\begin{center}
\includegraphics[width=0.8\columnwidth] {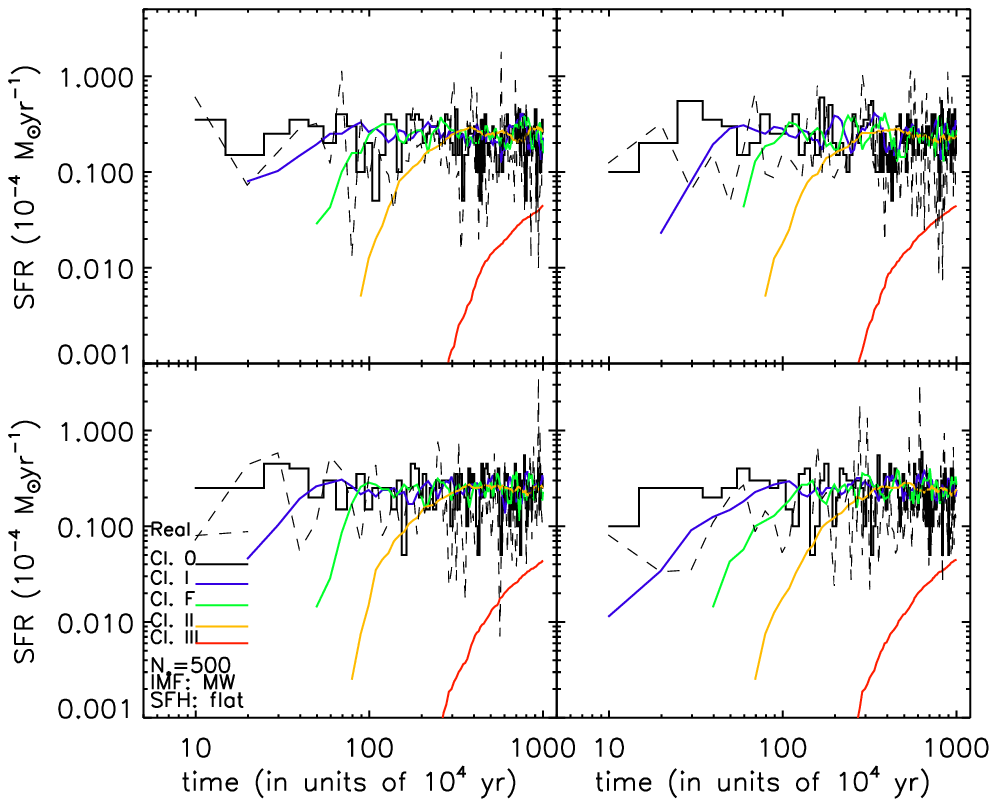}
\hspace{1.5cm}
\includegraphics[width=0.8\columnwidth] {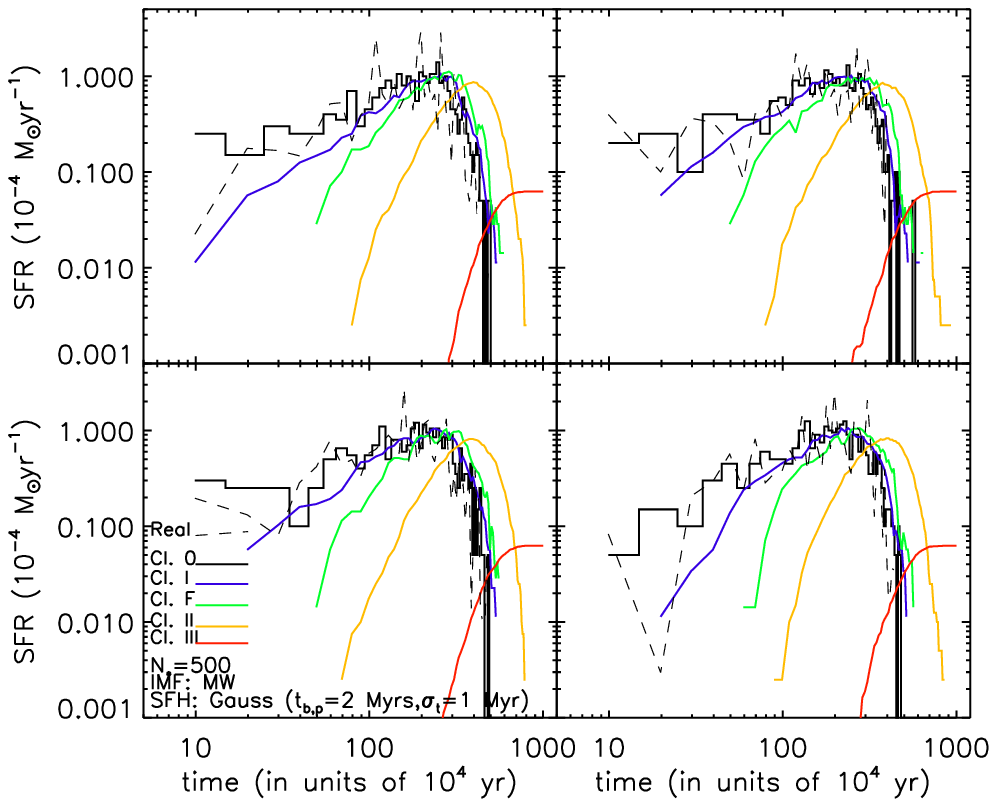}
\vspace{1cm}
\caption{Left: Four realizations of the time evolution of the SFR measured using the populations of YSOs found in different classes and their respective lifetimes. The dashed line displays the time evolution of the true SFR. Each cluster has $N_{*}=500$, a Milky Way-like IMF, and a constant SFH. Right: Similar to the left panel, but here each cluster has $N_{*}=500$, a Milky Way-like IMF, and a Gaussian-like SFH. The Gaussian's peak is at 2 Myr and the standard deviation is 1 Myr. } 
\label{fig5}
\end{center}
\end{figure*} 

\begin{figure*}
\begin{center}
\includegraphics[width=0.8\columnwidth] {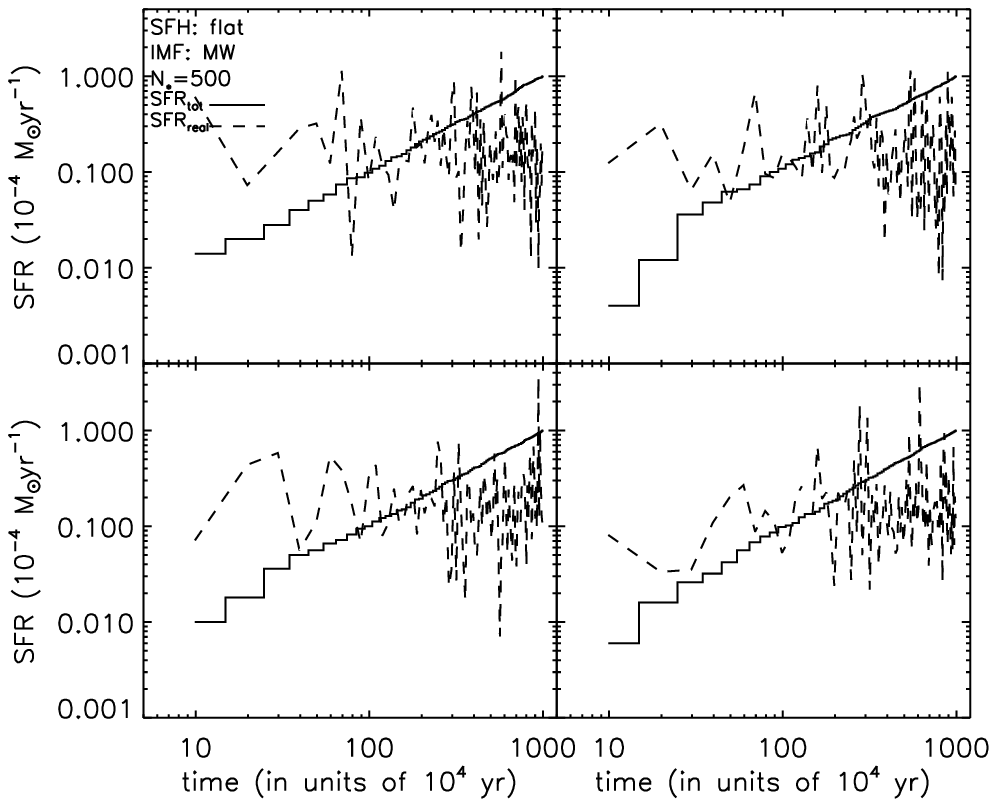}
\hspace{1.5cm}
\includegraphics[width=0.8\columnwidth] {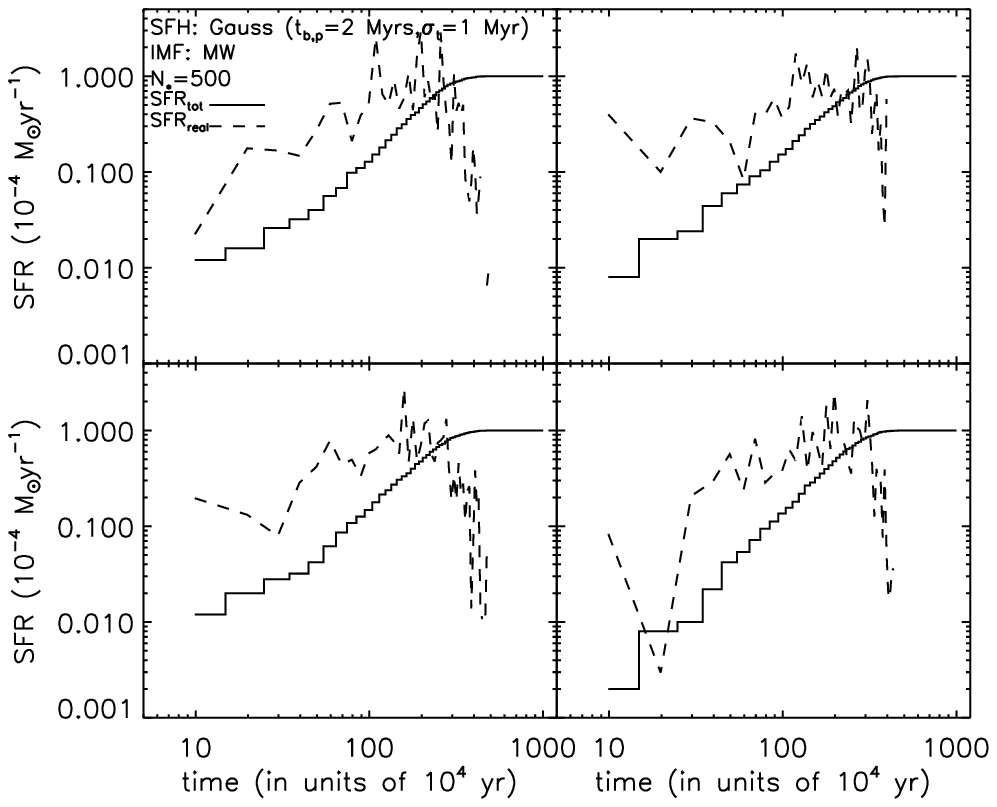}
\vspace{1cm}
\caption{Left: Four realizations of the time evolution of the true SFR (SFR$_{\rm real}$, dashed line) and the SFR measured using the total number of YSOs and an average lifetime for all Classes. Each clusters has $N_{*}=500$, a Milky Way-like IMF and a constant SFH. Right: Similar to the left panel, but here, each cluster has $N_{*}=500$, a Milky Way-like IMF, and a Gaussian-like SFH. The Gaussian's peak is at 2 Myr and the standard deviation is 1 Myr.} 
\label{fig6}
\end{center}
\end{figure*} 

\begin{figure}
\begin{center}
\includegraphics[width=0.95\columnwidth] {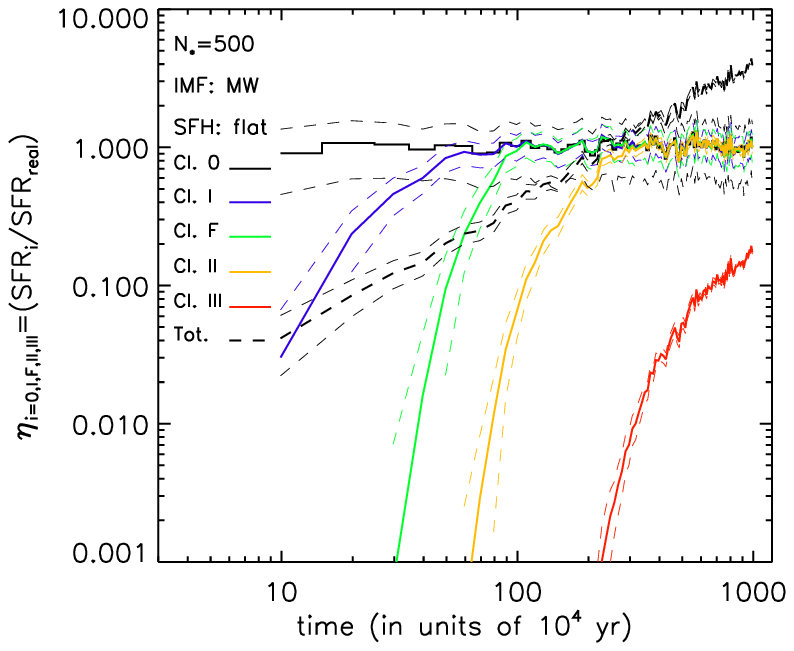}
\includegraphics[width=0.95\columnwidth] {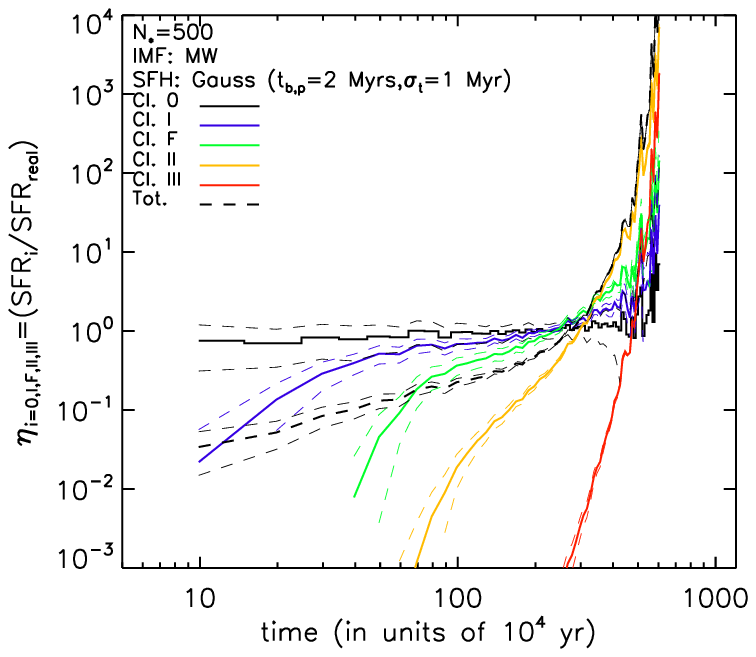}
\caption{Top: Time evolution of the ratio of the SFR measured using the populations of YSOs found in different classes (full lines) and the total population (dot-dashed line) and the real SFR that is generated in the models. Each line is an average over 250 clusters. Each cluster has N$_{*}=500$, a Milky Way-like IMF, and a flat SFH. Bottom: Similar to the top panel, but here each clusters has $N_{*}=500$, a Milky Way-like IMF, and a Gaussian-like SFH. The Gaussian's peak is at 2 Myr and the standard deviation is 1 Myr. In both panels, the dashed lines mark the 1$\sigma$ uncertainty levels.} 
\label{fig7}
\end{center}
\end{figure} 

When deriving the SFR using Eq.~\ref{eq1}, and in almost all observational studies, the assumption is made that the SFH in a star-forming region has been constant over the entire lifetime of the regions (Evans et al. 2009 and references therein). This assumption may be true in some molecular clouds, possibly in low-mass clouds where star formation proceeds at a slow pace, but is unlikely to be true for all clouds. The age distribution of stars in young clusters is not always a flat distribution as exemplified by the case of the Orion Nebula Cluster (ONC, Dib et al. 2013). In the ONC, the age distribution of stars is well reproduced by the sum of a linearly increasing SFH at old ages (up to 6 Myr) and an acceleration (i.e., a burst) of star formation at younger ages (see Figure 13 in Dib et al. 2013). Similar SFHs, consisting of a slow increase when clouds are in the process of assembling followed by a burst of star formation at later times as most of the gas sinks to the central regions of the clouds, are also observed in numerical simulations (e.g., V\'{a}zquez-Semadeni et al. 2017, Guszejnov et al. 2022). It has also been shown that such burst-like SFHs better describe the SFH of clusters such as the Orion Nebular cluster (Kounkel 2020; see also Megeath et al. 2022), and are necessary in order to reproduce the mass--luminosity relation of massive star-forming regions (Zhou et al. 2024c,d). In the present work, we test both assumptions. We consider cases where the SFH is constant and others where it varies over time. For cases with a constant SFH (labelled as constant or flat), we consider a fiducial case with a birthrate of $5\times10^{-5}$ stars yr$^{-1}$ (i.e., 500 stars form over a period of $10^{7}$ yr). For the time-dependent SFH, a variety of functional forms can be used to describe a slowly increasing SFH followed by a burst. For simplicity, we use a Gaussian function, for which we are able to vary the position of the peak and the width of the distribution. The SFH is described by a probability for the time of birth of the stars, $t_{\rm b}$, and in the case of a time-varying SFH this is given by

\begin{equation}
P(t_{b})  = \frac{1} {\sigma_{\rm t} \sqrt{2 \pi}} \exp\left(-\frac{1}{2} \left(\frac{t_{\rm b}-t_{\rm b,p}}{\sigma_{t}}\right)^{2}\right),
\label{eq4}
\end{equation}

\noindent where $t_{\rm b,p}$ is the position of the peak and $\sigma_{\rm t}$ the standard deviation. For a cluster whose stars form at a constant rate, their $t_{\rm b}$ are randomly drawn from a flat probability between $t=0$ and $t_{\rm end}$, where $t_{\rm end}$ is the time at which star formation ends. In the case of a time-dependent SFH, the $t_{\rm b}$ of stars are randomly drawn from the $P (t_{\rm b})$ function given in Eq.~\ref{eq4}. For the case with a time-dependent SFH, the fiducial parameters are $t_{\rm b,p}=2$ Myr and $\sigma_{\rm t}=1$ Myr. We take a $t_{\rm end}$ that is larger than any of the lifetimes associated with the protostellar classes and assume that $t_{\rm end}=10$ Myr. 

\section{Results}\label{res}

\subsection{The effect of the SFH}

First, we present results from two fiducial cases, one with a constant SFH and another with a Gaussian-like SFH. For these models, we consider a cluster that will end up containing $N_{*}=500$ stars sampled from a Milky Way-like single-star IMF (i.e., $\Gamma=1.35$, $\gamma=0.51$, and $M_{\rm ch}=0.35$ M$_{\odot}$). For each case, we generate 250 models. Figure \ref{fig1} displays four realizations of the Milky Way-like randomly sampled IMF. Figure \ref{fig2} and Fig.~\ref{fig3} each display four realizations of the birth times of protostars. For the case of the flat SFH displayed in Fig.~\ref{fig2}, the birth times of protostars are randomly selected with a uniform probability in the time span between $t=0$ and $t=t_{\rm end}$, whereas for the cases of the Gaussian-like SFH displayed in Fig.~\ref{fig3}, the birth times of protostars are randomly sampled using Eq.~\ref{eq4}. The next step is to assign to each protostar lifetimes that are associated with each protostellar class. As described above, these lifetimes are randomly drawn using the probability distribution functions described by Eq.~\ref{eq3}. As fiducial values, we adopt here those listed in \S.~\ref{classlifetimes}, namely $\bar{\tau}_{0}=0.1$ Myr and $\sigma_{\tau_{0}}=0.05$ Myr for Class 0, $\bar{\tau}_{\rm I}=0.44$ Myr and $\sigma_{\tau_{\rm I}}=0.22$ Myr for Class I, $\bar{\tau}_{\rm F}=0.35$ Myr and $\sigma_{\tau_{\rm F}}=0.175$ Myr for the F Class, $\bar{\tau}_{\rm II}=2$ Myr and $\sigma_{\tau_{\rm II}}=1$ Myr for Class II, and $\bar{\tau}_{\rm III}=4$ Myr and $\sigma_{\tau_{\rm III}}=2$ Myr for Class III. As an example, Fig.~\ref{fig3} displays the distribution functions of the protostellar lifetimes for all YSOs that form in one cluster. When randomly sampling each YSO lifetime from the Gaussian function shown in Eq.~\ref{eq3}, the randomly sampled values are drawn in the range of the $\pm 1\sigma$ uncertainty of each quantity. 

With the above input quantities, we track the number of YSOs found in each class as a function of time. Figure \ref{fig4} (left panel) displays four examples (out of 250 generated models) for the case with a flat SFH, a Milky-like IMF, and $N_{*}=500$ in each cluster. Figure~\ref{fig4} indicates that within the temporal fluctuations due to the random sampling, the total number of Class 0 protostars is constant over time. The number of Class I and Class F protostars starts to increase after $\approx 0.1$ and $0.4$ Myr, respectively, and both reach a plateau after $1$ Myr. The numbers of Class II and Class III PMS stars display the same behavior but at later times (i.e., after between two and several million years). The right panel of Fig.~\ref{fig4} displays the time evolution of the number of YSOs in each class for four cases with a Gaussian-like SFH (with the fiducial parameters). This figure clearly shows the peak in Class 0 protostars at $t=t_{\rm b,p}=2$ Myr and this peak shifts to later epochs for more evolved protostars. WE note that in these cases, the number of Class II and Class III PMS stars also declines at later times because most of the protostars were formed early at $t=2$ Myr and almost no new protostars were formed after $\approx 4$ Myr.

\begin{figure}
\begin{center}
\includegraphics[width=\columnwidth] {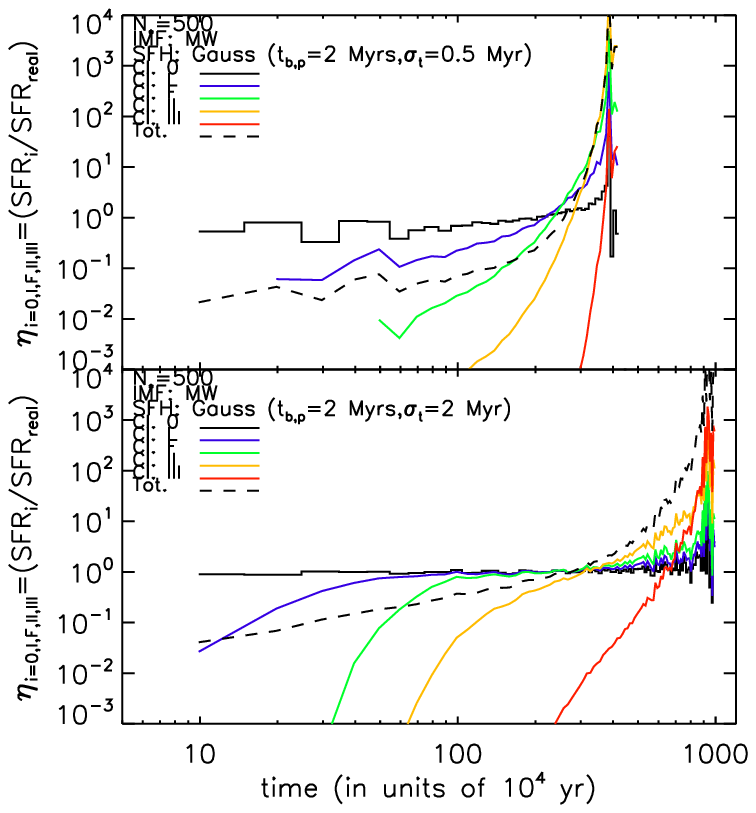}
\vspace{0.5cm}
\caption{Top: Time evolution of the ratio of the SFR measured using the populations of YSOs found in different classes (full lines) and the total population (dashed line) and the real SFR that is generated in the models. Each line is an average over 250 clusters. Each cluster has N$_{*}=500$, a Milky Way-like IMF, and a Gaussian-like SFH with the peak of the Gaussian located at $t_{\rm b,p}=2$ Myr and a standard deviation of $\sigma_{\rm t}=0.5$ Myr. Bottom: Similar to the top panel, but with $t_{\rm b,p}=2$ Myr and $\sigma_{\rm t}=2$ Myr. For improved clarity, the $1\sigma$ uncertainties are not shown.} 
\label{fig8}
\end{center}
\end{figure} 

Using the number of YSOs found in each class and at each epoch, as well as the mean values of their lifetimes (i.e., those suggested by Evans et al. 2009), we can now apply Eq.~\ref{eq1} and calculate the corresponding SFRs. We also measure the true SFR in each model. For this, we take a time step of $\Delta t=10^{4}$ yr. This guaranties that the smallest protostellar lifetimes are well resolved. We then simply calculated the true SFR as the total mass of protostars that form within each time step:

\begin{equation}
{\rm SFR_{real}}(t+\Delta t/2)=\frac{\sum_{j} M_{*,j}}{\Delta t}, 
\label{eq5}
\end{equation}

\noindent where $j$ runs over all stars that form between $t$ and $t+\Delta t$. Figure~\ref{fig5} (left) displays the time evolution of the SFR for four cases where the SFH is flat. We recall that this is usually the assumption that is made in the observations when deriving the SFR. Figure \ref{fig6} (left panel) displays the true SFR measured directly from the models (dashed lines) and an SFR estimate using the total number of YSOs (full line). For the latter, a characteristic YSO lifetime of 2.5 Myr was used (Pokhrel et al. 2020). The left panels of Fig.~\ref{fig5} and Fig.~\ref{fig6} show that, while the SFR estimates measured using a single class converge to the real value, they do so with a delay that increases for more evolved YSOs. This implies that for cases where the SFH is constant, measurements of the SFR will only be reliable if only Class 0 and I protostars are considered in young star forming regions and Class II PMS stars in older regions (i.e., ages $\gtrsim 2$ Myr). Figure \ref{fig6} (left panel) additionally shows that using the total number of YSOs provides a very poor approximation of the real SFR; it underestimates the true SFR by a factor of between $\approx 10$ and $100$ in young star-forming regions and overestimates the true SFR by a factor of a few to several for older regions. The right panels in Fig.~\ref{fig5} and Fig.~\ref{fig6} display the same SFRs calculated for cases where the SFH is a Gaussian-like function with the fiducial values of the parameters, $t_{\rm p,b}=2$ Myr and $\sigma_{\rm t}=1$ Myr. The same patterns that are observed with a constant SFH are also prevalent here, with the additional effect that at advanced epochs (i.e., ages $\gtrsim 3.5$ Myr), the estimates of the SFR that are  either based on the Class II and Class III PMS stars or on the total number of YSOs overestimate the true SFRs by several orders of magnitude. This is due to the fact that while star formation has ceased, Class II and Class III PMS stars are long lived and their numbers bias the measurement of the SFR. 

In order to get a clearer picture of the offset between the SFRs estimated using the YSO populations and the true SFR, we show the ratio between these quantities  in Fig.~\ref{fig7}. In both panels of this figure, the ratios were obtained by averaging over the 250 realizations (i.e., 250 clusters), each with $N_{*}=500$ and a Milky Way-like IMF. The top panel corresponds to cases with a flat SFH and the bottom panel to cases with the Gaussian-like SFH, with all clusters in the latter case having $t_{\rm b,p}=2$ Myr and $\sigma_{\rm t}=1$ Myr. The light dashed line in both panels corresponds to the 1$\sigma$ Poisson uncertainty on each measurement. Ideally, for these measurements to be accurate, the offset parameter $\eta_{i=\rm 0,I,F,II,III}$=SFR$_{i}$/SFR$_{\rm real}$ would have to be close to the unity. However, this is far from being the case, with the worst departure from unity corresponding to the time-dependent, Gaussian-like SFH, especially around and after the peak of star formation at $\approx 2$ Myr. For these time-dependent models of the SFH, we expect the level of departure from unity to depend on the duration of the burst (i.e., the width of the Gaussian). We therefore ran additional models, each with 250 clusters with Gaussian-like SFHs. As our fiducial case has a value of $\sigma_{\rm t}=1$ Myr, we considered additional cases where $\sigma_{\rm t}=0.5$ Myr (a narrow, high-amplitude burst) and $\sigma_{\rm t}=2$ Myr (an extended, low-amplitude burst). The results for these models are displayed in Fig.~\ref{fig8} for the narrow burst (top panel) and the extended burst (bottom panel). Qualitatively, both cases display the same features and are similar to the fiducial case displayed in Fig.~\ref{fig7} (top panel). However, a noticeable difference for the narrow-burst case (i.e., case with $\sigma_{\rm t}=0.5$ Myr) is that all SFR measurements using the different classes of YSOs or their total number, with the exception of the one based on the Class 0 population, fail to reproduce the true SFR.   

The conclusion that can be drawn from Fig.~\ref{fig7} and Fig.~\ref{fig8} is that the only reliable estimate of the SFR at all epochs is the one based on the population of Class 0 protostars. The latter measurement is always accurate, irrespective of the SFH. Measurements of the SFR based on the population of more evolved YSOs can provide reliable estimates of the SFR in cases where the SFH is constant or displays a weak burst, and this is only valid when the star-forming regions are themselves more evolved (i.e., with ages of 2 Myr or older, and harboring a significant fraction of evolved YSOs). In all circumstances, estimates of the SFR that rely on the total number of YSOs and a characteristic YSO lifetime (in this work, $2.5$ Myr) are not accurate and this is true irrespective of the SFH and the considered age of the region. Finally, we should mention that we ran models with lower and higher numbers of YSOs, namely $N_{*}=250$ and $1000$, respectively, while keeping $t_{\rm end}=10$ Myr in all models. Models with lower or higher numbers of protostars display the same effects as those observed in models with $N_{*}=500$ YSOs, but with an increased and decreased level of temporal fluctuation, respectively (see Appendix \ref{appa}).

\subsection{The effect of the protostellar mass function}\label{effetimf}

\begin{figure}
\begin{center}
\includegraphics[width=0.9\columnwidth] {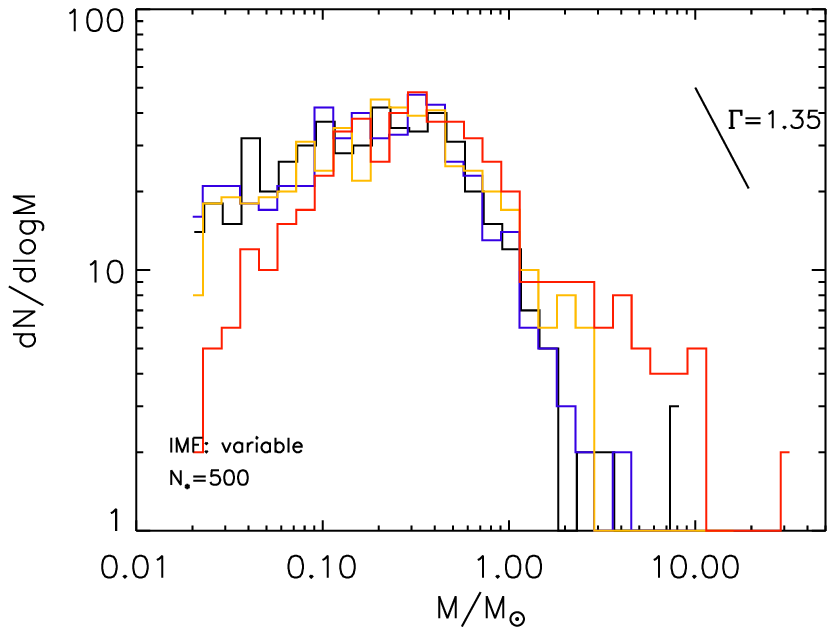}
\caption{Four realizations of the variable protostellar mass function. The masses of $N_{*}=500$ protostars are randomly drawn from a TPL function in the stellar mass range of 0.02 to 50 M$_{\odot}$. For each realization, the parameters that characterize the shape of the TPL mass function are randomly drawn from Gaussian distribution functions (see text for more details).}
\label{fig9}
\end{center}
\end{figure} 

\begin{figure}
\begin{center}
\includegraphics[width=0.9\columnwidth] {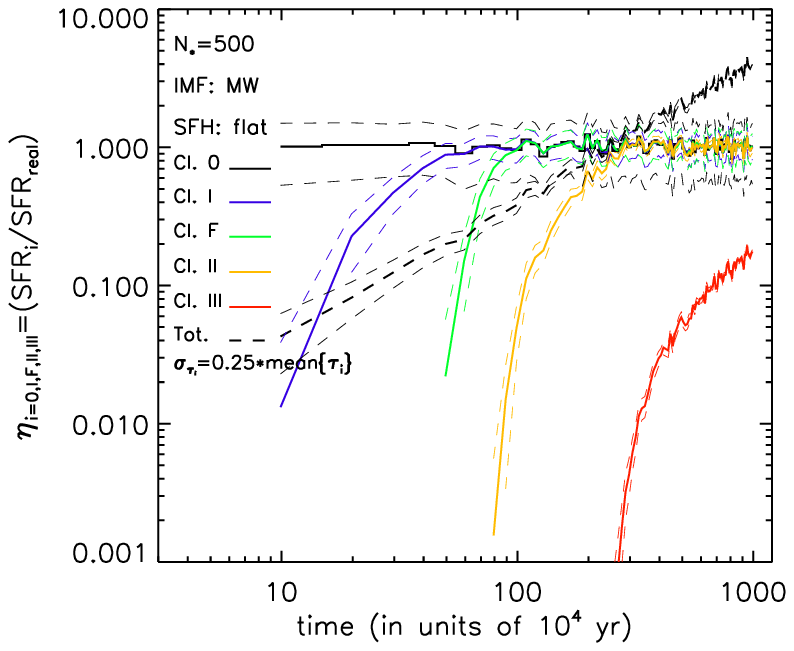}
\vspace{1cm}
\caption{Same as the top panel of Fig.~\ref{fig7} but in this case the parameters that describe the shape of the IMF for each of the 250 clusters are randomly drawn from Gaussian distribution functions (see text for more detail).  For more clarity, the $1\sigma$ uncertainties are not shown.}
\label{fig10}
\end{center}
\end{figure}

Recent studies suggest a non-negligible degree of variation in the parameters that characterize the shape of the IMF of young clusters in the Milky Way and in M31 (Dib 2014; Weisz et al. 2015; Lim et al. 2015; Dib et al. 2017, Dib 2023b). Dib et al. (2017) found that an intrinsic scatter in the IMF parameters is necessary in order to match the observed fraction of single O stars in young  Galactic clusters (cluster ages $\lesssim 12$ Myr). For a TPL function representing the system IMF, Dib et al. (2017) found that the standard deviation in $\Gamma$, $\gamma$, and $M_{\rm ch}$ is $\approx 0.6$, $0.25$, and $0.27$ M$_{\odot}$. Here, as we are dealing with a single-star protostellar mass function, we model the distribution of each of the mass function parameters with a Gaussian centered around the Galactic values ($\Gamma=1.35$, $\gamma=0.51$, and $M_{\rm ch}=0.35$ M$_{\odot}$) and with a standard deviation on each parameter that is $50\%$ of its mean value, namely $\sigma_{\Gamma}=0.6$, $\sigma_{\gamma}=0.25$, and $\sigma_{M_{\rm ch}}=0.17$ M$_{\odot}$. Similar to the previous calculations, we generate a sample of 250 clusters, each with $N_{*}=500,$ and where the parameters of the IMF of each cluster are randomly drawn from these Gaussian distributions. The IMF parameters are randomly drawn from their Gaussian distribution within the $\pm 1\sigma$ uncertainty range. Figure \ref{fig9} displays the cases for four clusters with such a variable IMF. We then follow the same procedure as above and derive the SFRs based on the different YSO classes and an SFR based on the total YSO population. Adopting the case of a flat SFH, Fig.~\ref{fig10} displays the time evolution of these SFRs normalized by the true SFR. Comparing this figure with Fig.~\ref{fig7} (top panel), we conclude that a variable IMF has little effect on the derived SFRs. The same features observed in the top panel of Fig.~\ref{fig7} and where the IMF of clusters are randomly drawn from a Milky Way-like IMF are seen in Fig.~\ref{fig10}; that is, we see a faithful reproduction of the true SFR (SFR$_{\rm real}$) by the SFR based on the Class 0 population, and a reproduction of the SFR$_{\rm real}$ by the SFRs based on the more evolved populations (Class I and beyond) at more advanced epochs. The SFR derived using the total YSO population remains a poor approximation of the true SFR, as it severely underestimates SFR$_{\rm real}$ at early epochs and overestimates it at later epochs.  

\section{Constraining the SFH using protostellar counts and SFR estimates}\label{correct}

\begin{figure}
\begin{center}
\includegraphics[width=0.8\columnwidth] {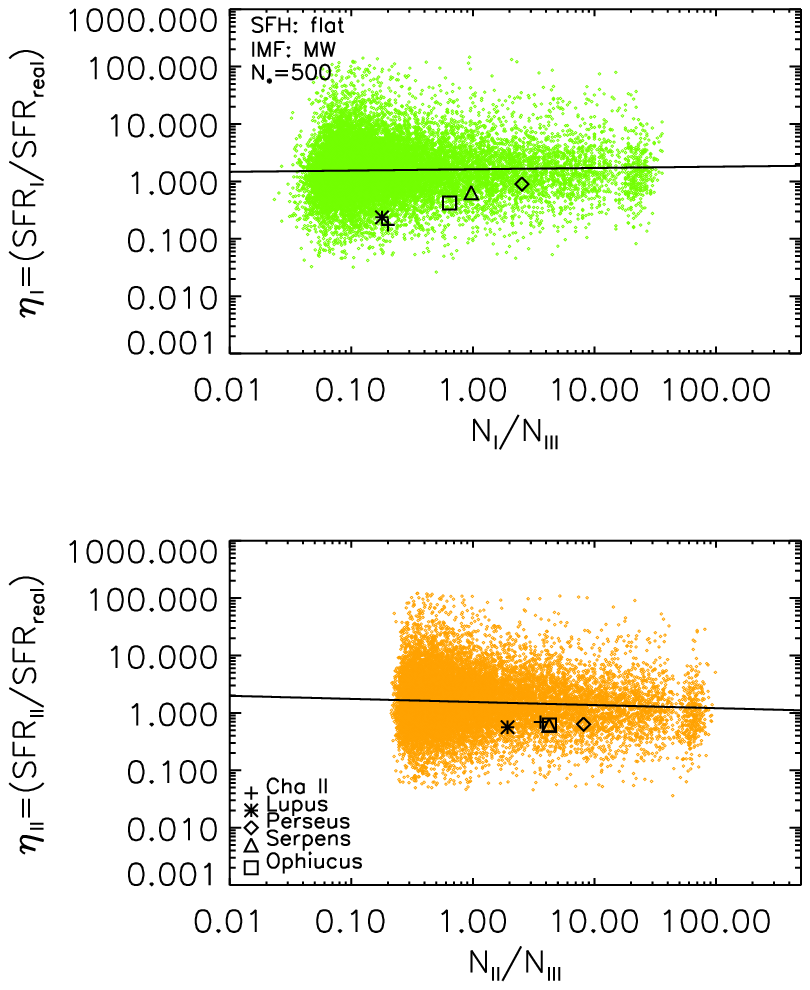}
\caption{Top panel: Offset between the SFR derived from the Class I population and the true SFR as a function of the ratio between the numbers of Class I and Class III PMS stars. The figure contains data from 250 clusters, each containing $N_{*}=500,$ and for each cluster, it includes measurements at 100 epochs. The SFH is flat and the IMF resembles that of the Milky Way field. Bottom panel: Similar to the top panel but showing the offset between the SFR derived from the Class II population and the true SFR as a function of the ratio of the numbers of Class II and Class III PMS stars.}
\label{fig11}
\end{center}
\end{figure} 

\begin{figure}
\begin{center}
\includegraphics[width=0.8\columnwidth] {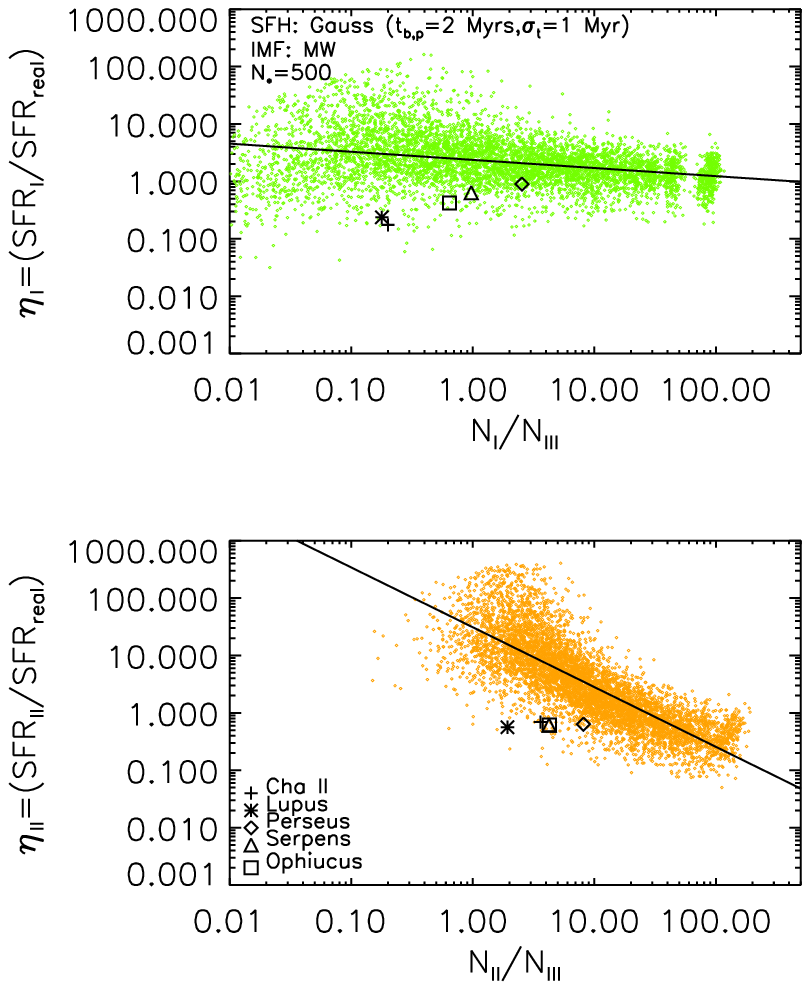}
\caption{Same as Fig.~\ref{fig11} but in this case the SFH is a time-dependent Gaussian-like function with the peak of the Gaussian located at $2$ Myr and a standard deviation of $\sigma_{\rm t}=1$ Myr.}
\label{fig12}
\end{center}
\end{figure} 

In \S.~\ref{res}, we argue that the SFR derived using the Class 0 YSO count provides a reliable estimate of the true SFR in star-forming regions, irrespective of the SFH. However, as exemplified by the models with $N_{*}=500$ stars, the number of Class 0 YSOs remains quite modest in most, if not all, epochs. Furthermore, Class 0 YSOs are sometimes difficult to detect as they are deeply embedded in the clouds (Evans et al. 2009). In contrast, the numbers of evolved YSOs is much larger. There are approximately 20 times more Class II YSOs than Class 0 YSOs. For this reason, using Class II objects to derive the SFR can be an appealing option in order to make use of the larger number of  Class II objects. A caveat here is that the offset between the Class II-derived SFR (SFR$_{\rm II}$) and the true SFR depends on the SFH. As shown in \S.~\ref{res}, for the case of a flat SFH, the offset between SFR$_{\rm II}$ and SFR$_{\rm real}$ exists only at early epochs ($\lesssim$ 2 Myr from the start of star formation). For the case of a burst-like SFH, the situation is more complex, and an offset between SFR$_{\rm II}$ and SFR$_{\rm real}$ exists at all times. It is possible to reverse the arguments and assumes that the observed SFR$_{\rm II}$ corresponds to the SFR$_{\rm real}$ and verify whether this assumption holds for the predictions of a constant SFH or a burst-like one. Figure \ref{fig11} displays the Class I and Class II SFR offsets plotted against the ratios $\left(N_{\rm I}/N_{\rm III}\right)$ and $\left(N_{\rm II}/N_{\rm III}\right)$, respectively, for the case with a flat SFH. Figure \ref{fig12} is similar to Fig.~\ref{fig11} but for the fiducial case with a Gaussian-like SFH. The dependencies between the offsets $\eta_{\rm I}$ and $\eta_{II}$ and the number ratios are noticeably different between these cases with a different SFH. 

We can test the usefulness of these relations in discriminating between SFHs by including observational data. We use the data presented in Tables 3 and 5 of Evans et al. (2009) for five nearby molecular clouds, namely Cha II, Lupus, Perseus, Serpens, and Ophiucus. Table 5 of Evans et al. (2009) lists the numbers of YSOs in Class I, the F Class, Class II, and Class III. As the numbers of Class II YSOs are larger than those in all other classes, this indicates that star formation started at least $\approx 1$ Myr ago, which is roughly equal to $\bar{\tau}_{0}+\bar{\tau}_{\rm I}+\bar{\tau}_{F}$ plus a fraction of the Class II lifetime. Table 3 of Evans et al. (2009) also lists SFR estimates for these clouds. These SFR measurements are derived using the Class II YSO populations and were presented in separate works, but are all based on c2d observations. Using the number counts of Class I and Class II YSOs, we use Eq.~\ref{eq1} to calculate SFR$_{\rm I}$ and SFR$_{\rm II}$ and consequently $\eta_{\rm I}$ and $\eta_{\rm II}$. If  the SFR values quoted for these clouds are the true values, the ratio $\eta_{\rm II}$ for these clouds should coincide with the prediction of the model with a constant SFR. The results shown in  Fig.~\ref{fig11} (bottom panel) confirm that this is indeed the case. Furthermore, the upper panel of Fig.~\ref{fig11} shows that the $\eta_{\rm I}$ versus $(N_{\rm I}/N_{\rm III})$ relation for these five clouds is also reasonably well matched by the model with a constant SFH. In contrast, the model with a burst-like SFH does not overlap with the observational data, as can be seen in the lower and upper panels of Fig.~\ref{fig12}. The conclusion that can be drawn from Figs.~\ref{fig11} and~\ref{fig12} is that the SFH in these five nearby and low- to intermediate-mass clouds (from $\approx 400$ to $\approx 5000$ M$_{\odot}$; Table 1 in Evans et al. 2009 and references therein) is relatively constant. It would be interesting to include data points for more massive star-forming regions and regions that are located further away in the Galaxy.     

Below, we list the fits to the $\eta_{\rm I}-\left(N_{\rm I}/N_{\rm III} \right)$ and $\eta_{\rm II}-\left(N_{\rm II}/N_{\rm III}\right)$ relations. For the flat SFH case, we find:

\begin{equation}
\begin{aligned}
\eta_{\rm I}  = \frac{\rm {SFR_{I}}} {\rm {SFR_{\rm real}}}  &=1.66(\pm 0.017) \left(\frac{N_{\rm I}}  {N_{\rm III}} \right)^{0.026 \pm 0.005} \\
\eta_{\rm II} = \frac{\rm {SFR_{II}}} {\rm {SFR_{\rm real}}} &=1.59(\pm0.011) \left(\frac{N_{\rm II}} {N_{\rm III}}\right)^{-0.063 \pm 0.006},
\label{eq6} 
\end{aligned}
\end{equation}

and for the time-dependent, Gaussian-like SFH, we find:

\begin{equation}
\begin{aligned}
\eta_{\rm I} = \frac{\rm {SFR_{I}}} { \rm {SFR_{\rm real}}}  &= 2.33(\pm 0.029) \left(\frac{N_{\rm I}}  {N_{\rm III}} \right)^{-0.13 \pm 0.005} \\
\eta_{\rm II} = \frac{\rm {SFR_{II}}} {\rm {SFR_{\rm real}}} &=27.25 (\pm 0.68) \left(\frac{N_{\rm II}} {N_{\rm III}}\right)^{-0.98 \pm 0.01}.
\label{eq7} 
\end{aligned}
\end{equation}

 As pointed out in \S.~\ref{classlifetimes}, the Class III YSOs observed by Spitzer for these nearby star-forming regions are incomplete and Eqs.~\ref{eq6} and \ref{eq7} should be revised with more complete class III censuses. More complete censuses of Class III YSOs could be provided by X-ray observations, as shown by Preibisch et al. (2005) in the Orion Nebular Cluster. However, we note that more accurately accounting for Class III YSOs will reinforce our conclusions concerning the comparison of our models with the observations. Accounting for a higher number of missing Class III YSOs would cause the observational number ratios $\left(N_{\rm I}/N_{\rm III}\right)$ and $\left(N_{\rm II}/N_{\rm III}\right)$ in Figs.~\ref{fig11} and \ref{fig12} to move to the left. This would not affect the agreement between the flat SFH models and the observations, but it would cause burst-like models to deviate further from the observations. While it is not yet possible to include a more massive region, such as the entire Orion cloud, because the census of Class II YSOs is incomplete (Furlan et al. (2016)\footnote{Megeath et al. (2012,2016) performed a comprehensive survey of the YSO population in the Orion clouds. However the YSOs in these works were categorized in the broad category of protostars and PMS stars and for this reason, the Orion data points could not be included in Figs.~\ref{fig11} and \ref{fig12}}. Furlan et al. (2016) found that the number of Class 0 YSOs in Orion ($N_{0}=92$). Using Eq.~\ref{eq1}, we conclude that the true SFR for Orion is $4.6\times10^{-4}$ M$_{\odot}$ yr$^{-1}$. This value sits in the middle between the SFRs measured by Lada et al. (2010) for the Orion A and Orion B clouds.   
 
\section{Conclusions}\label{conc}

In this work, we explore the circumstances  under which counting YSOs in star-forming regions can yield reliable estimates of the SFR. To this end, we developed a Monte Carlo model in which the masses, the birth times of the YSOs, and the lifetimes associated with the different YSO classes (Classes 0, I, F, II, and III) are all randomly drawn from
distribution functions that describe each of these quantities. The masses of the protostars are randomly drawn from an IMF that can be either similar to that of the Milky Way field or variable in the range of variations observed among young clusters in the Galaxy. The birth times of protostars are either drawn from a flat distribution function (i.e., constant SFH), or from a time-dependent, burst-like function. Finally, the lifetimes of the YSOs associated with each protostellar class are randomly drawn from Gaussian distribution functions centered around those suggested by the observations (Evans et al. 2009). Using these prescriptions, we follow the time evolution of the number of YSOs in different evolutionary classes and calculate the corresponding SFR at every epoch. We find that the SFRs derived using the Class 0 population reproduce the true SFR, and this conclusion is valid irrespective of the shape of the SFH. For a constant SFH, the SFRs derived using the more evolved populations of YSOs (Class I, Class F, Class II, and Class III) reproduce the real SFR at later epochs. For example, for an SFR estimate that is based on the Class II population and lifetimes, the real SFR is reproduced after $\gtrsim 2$ Myr from the beginning of star formation. For a time-dependent burst-like SFH, all SFR estimates based on the number counts of the more evolved populations (all classes but Class 0) fail to reproduce the true SFR. Also, we find that SFR estimates that are based on the total number of YSOs associated with a characteristic lifetime for all populations fail to reproduce the real SFR, irrespective of the shape of the SFH. Furthermore, we show that all of  these conclusions are independent of the shape of the stellar IMF. 

The synthetic models presented in this work can help shed light on the SFH of observed star-forming regions. We show that the models with a constant SFH make different predictions for the SFR offsets versus the number ratios of YSOs from those with a burst-like SFH. For five, nearby, low-mass star-forming regions that constitute the Evans et al. 2009 sample (i.e, the Cha II, Lupus, Perseus, Serpens, and Ophiucus clouds), we show that the relations between the SFR offset parameters $\eta_{\rm I}={\rm SFR}_{\rm I}$ and $\eta_{\rm II}={\rm SFR}_{\rm II}$ of the clouds and the number ratios $N_{\rm I}/N_{\rm III}$ and $N_{\rm II}/N_{\rm III}$, respectively, are matched by models with a constant SFR. A comparison between our models and a larger number of star-formation regions, such as those that would be observed by the ABYSS project (Kounkel et al. 2023) and with the James Webb Science Telescope (e.g., Lenki\'{c} et al. 2024; Peltonen et al. 2024) will enable us to distinguish between the SFHs of star-forming clouds as a function of some of their fundamental properties, such as mass and surface density. In this work, we have not accounted for the effects of mass accretion, which induces variability in the luminosity of YSOs and may lead to their misclassification. We believe however that such effects will not change our conclusion, as only a small fraction of the YSO population shows signs of variability (Mairs et al. 2017).
 
Deriving accurate SFRs is crucial for the interpretation of the scaling laws of star formation on individual molecular cloud scales and for testing both theories and numerical simulations of star formation (e.g., Hennebelle \& Chabrier 2011; Dib 2011; Padoan \& Nordlund 2011; Federrath \& Klessen 2012; Burkhart 2018; Eden \& Teyssier 2024). In particular, the SFR--surface gas density ($\Sigma_{\rm g}$) relation on local scales is observed to be steeper than the one for entire galaxies or the one derived on approximately kiloparsec scales within galaxies (e.g., Heiderman et al. 2010; Hony et al. 2015; Pokhrel et al. 2020,2021). One interpretation of this steep relation between $\Sigma_{\rm g}$ and the SFR is that the higher SFR observed in individual regions is due to the selection bias introduced by studying a region that is actively forming stars, while the galaxy-averaged values include regions that are forming stars at different rates (Kruijssen \& Longmore 2014). Another interpretation is that the steep $\Sigma_{\rm g}$--SFR relation for individual clouds is the reflection of an evolutionary sequence in which the SFRs of contracting clouds increase much faster than their mean densities or mean surface densities and this leads to a steep supra-linear $\Sigma_{\rm g}$--SFR relation (Zamora-Avil\'{e}s et al. 2012). Further work is needed in order to disentangle these effects from the offsets in the SFR estimates uncovered in the present work, which are solely due to biases introduced by the star counting method.  

\begin{acknowledgements}

We thank the anonymous referee for useful comments and suggestions that helped clarify aspects of this work. SC acknowledges funding from the State Research Agency (AEI-MCINN) of the Spanish Ministry of Science and Innovation under the grant `Thick discs, relics of the infancy of galaxies' with reference PID2020-113213GA-I00. M.A.L.L. acknowledge support from the Ram\'{o}n y Cajal program funded by the Spanish Government (reference RYC2020-029354-I), and from the Spanish grant PID2021-123417OB-I00.
 
\end{acknowledgements}

{}

\begin{appendix}
\section{The effect of the number of stars}\label{appa}

\begin{figure}
\begin{center}
\includegraphics[width=0.9\columnwidth] {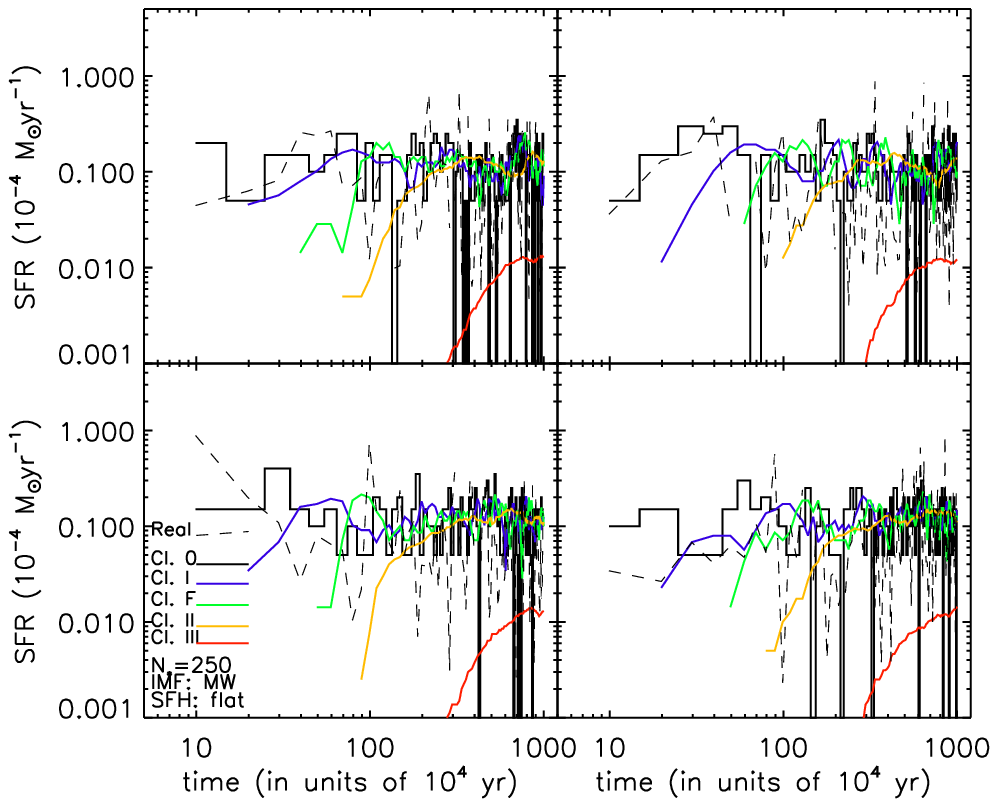}\\
\vspace{1.5cm}
\includegraphics[width=0.9\columnwidth] {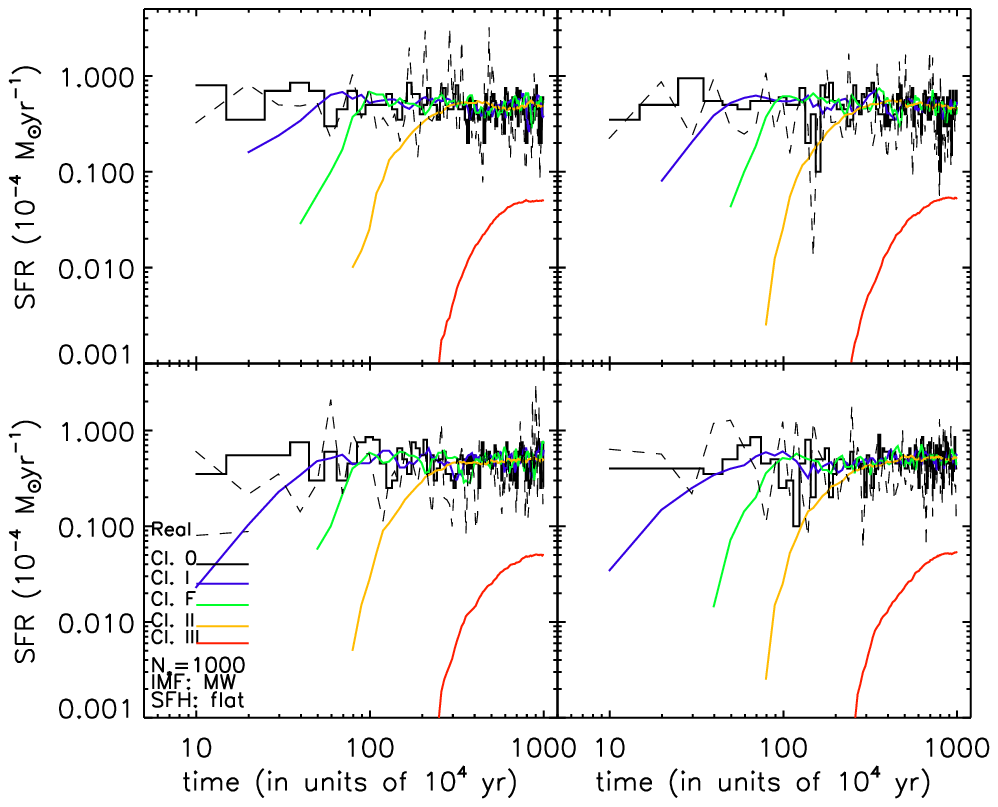}
\vspace{1cm}
\caption{top: Four realizations of the time evolution of the SFR measured using the populations of YSOs found in different classes and their respective lifetimes. The dashed line displays the time evolution of the true SFR. Each cluster has $N_{*}=250$, a Milky Way-like IMF and a constant SFH. bottom: Similar to the top panel, but here, each cluster has $N_{*}=1000$. } 
\label{figapp1}
\end{center}
\end{figure} 

\begin{figure*}
\begin{center}
\includegraphics[width=0.8\columnwidth] {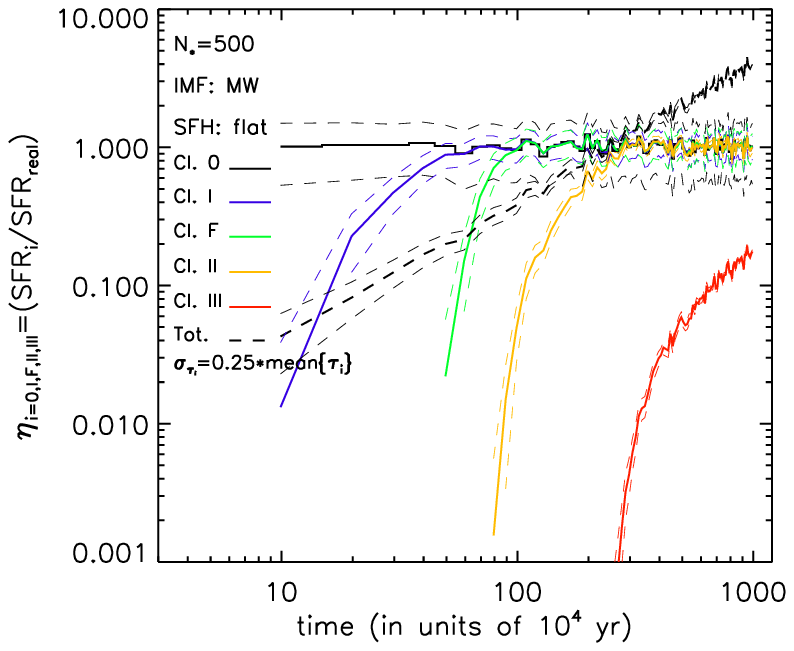}
\hspace{1cm}
\includegraphics[width=0.8\columnwidth] {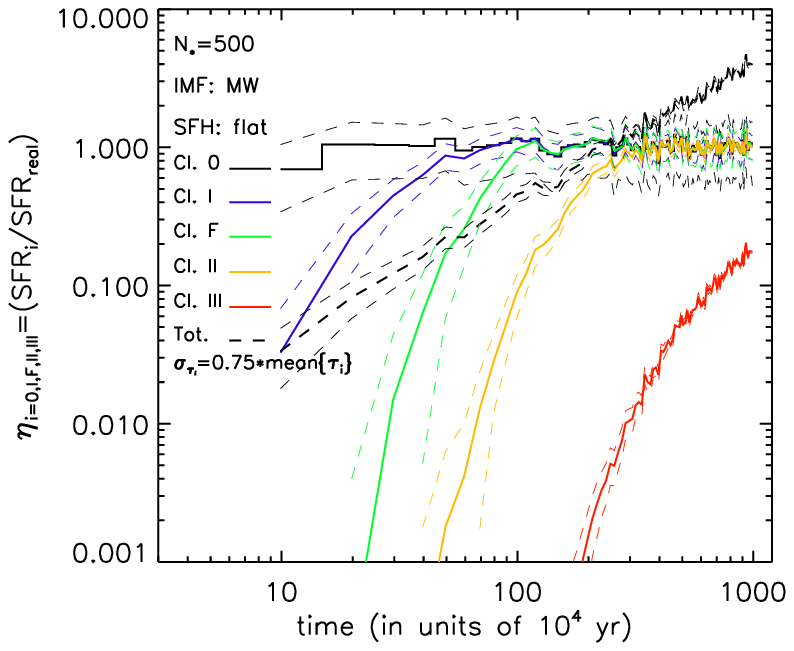}\\
\hspace{1cm}
\includegraphics[width=0.8\columnwidth] {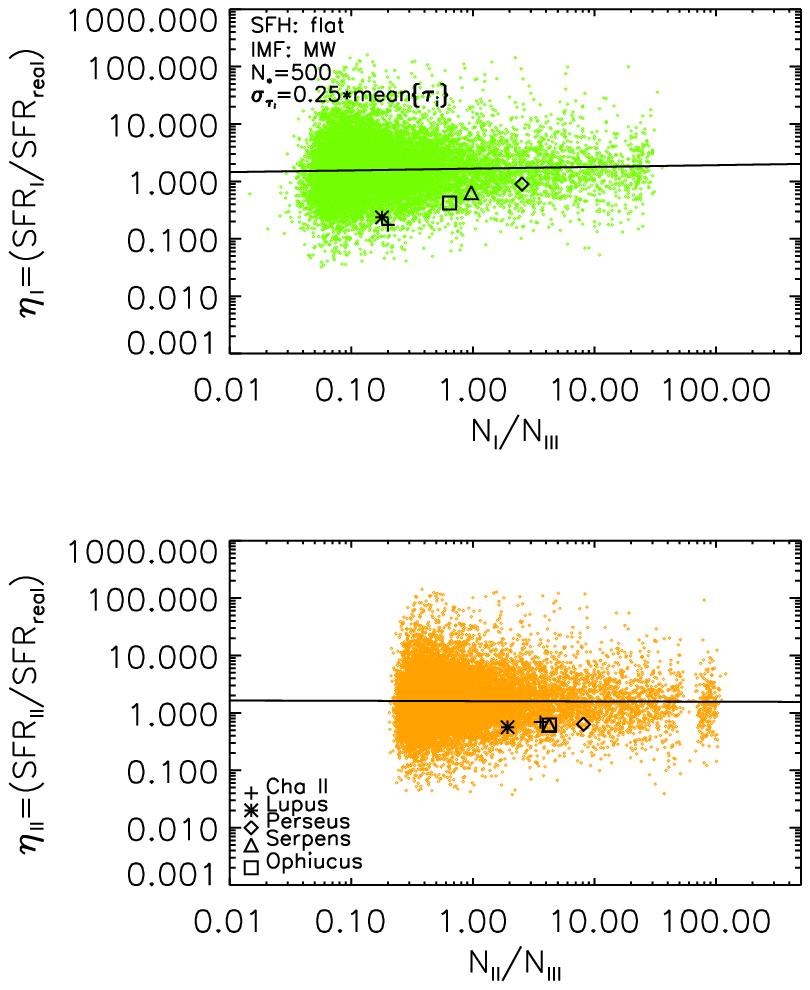}
\hspace{1cm}
\includegraphics[width=0.8\columnwidth] {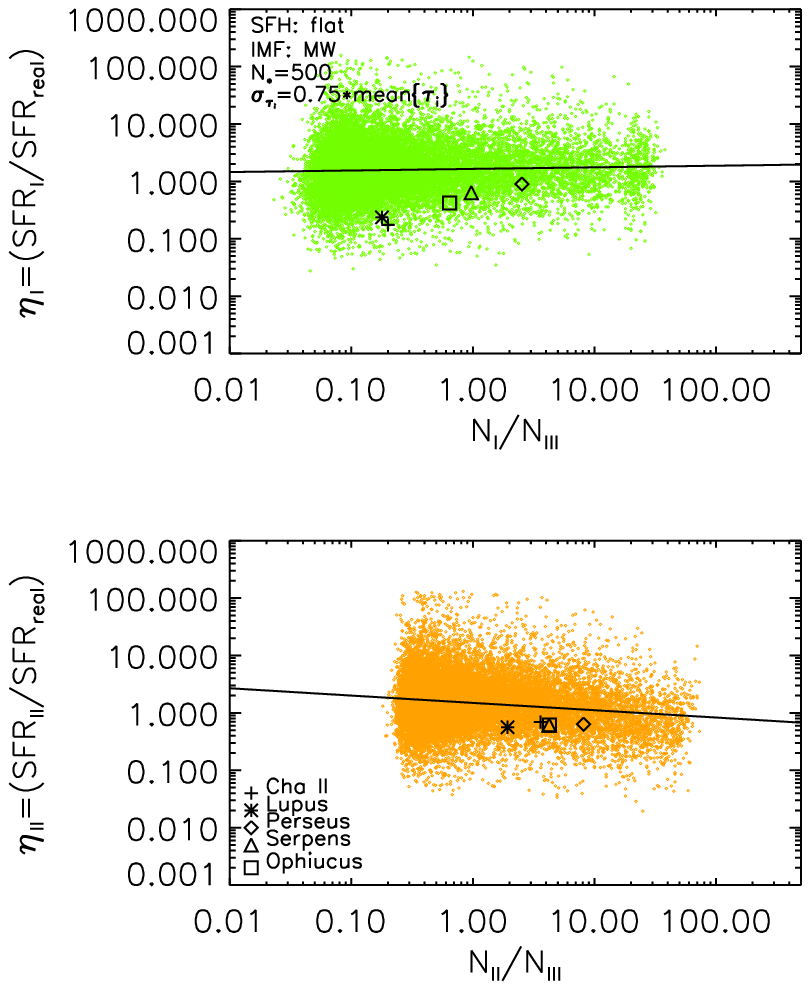}\\

\caption{Top row: same as Fig.~\ref{fig7} (top panel) but for cases where the width of the Gaussian distribution of the YSO lifetimes are taken as 0.25 the mean value (left) and 0.75 the mean value (right). Bottom row: Similar to Fig.~\ref{fig11} but for cases where the width of the Gaussian distribution of the YSO lifetimes are taken as 0.25 the mean value (left) and 0.75 the mean value (right).} 
\label{figapp2}
\end{center}
\end{figure*}

In the paper, we have chosen $N_{*}=500$ as the fiducial number of stars that form in each cluster, both for cases with a constant SFH and a time-dependent, burst-like SFH. Here we show examples of the calculated SFR in each protostellar Class for clusters with $N_{*}=250$ and $N_{*}=1000$. Fig.~\ref{figapp1} displays four realizations with $N_{*}=250$ and $N_{*}=1000$ for cases with a constant SFR and a Milky Way-like IMF. Decreasing (or increasing) the numbers of YSOs that form in the cluster by a certain factor (here a factor of 2) decreases (increases) the SFR by the same factor but a smaller $N_{*}$ (or larger $N_{*}$) increases (decreases) the temporal fluctuations in the SFR. 

\section{The effect of the YSOs' lifetime distribution functions}

In the main text, we have adopted a width of the Gaussian distributions for each YSO lifetime that is 0.5 times the mean lifetime value. Here, we explore how the width of such distribution affects our results. We consider two cases where the width of the lifetimes' Gaussian distributions is 0.25 and .0.75 their mean values. In both cases, we use a flat SFH, $N_{*}=500$, and adopt a Milky Way-like IMF. The results for the offset parameters are displayed in the top row. These figures show that when the distribution of YSO lifetimes are narrower, the saturated level of the offset ratio is reached earlier. This is simply due to the fact that narrower (wider) distributions imply that there are less (more) YSO that transition from a given class to the next one with lifetimes that are larger than the mean value of the corresponding lifetime. The effects of the YSOs lifetime distribution is reflected in the variations that can be observed in the offset vs. number ratios diagram. A comparison of the lower panels in Fig.~\ref{figapp2} along with Fig.~\ref{fig11} shows that these diagrams are affected by the level of variations in YSO lifetimes for each class, even though at this stage, it is difficult to distinguish which set of models fits the data best given the limited number of observational data.

\end{appendix}
 
\label{lastpage}

\end{document}